%% file: ms.tex
\documentclass[aip,prb,amsmath,amssymb, reprint]{revtex4-2} 

\usepackage{amsmath}  
\usepackage{amsfonts} 
\usepackage{graphicx} 
\usepackage{subcaption}
\usepackage{ragged2e}

\begin{document}


\title{Body on the hemispherical hill with friction \underline{and drag force}. A classical problem becoming more realistic with exact and perturbative approaches.}

\author{Gniewoj Michalewski}
\email{gniewojm@gmail.com} 
\affiliation{Polish Children's Fund, Warsaw, Poland}
\affiliation{Inter-faculty Individual Studies in Mathematics and Natural Sciences, University of Warsaw, Warsaw, Poland}
\date{25 July 2023}

\begin{abstract}
I discuss the influence of adding the air resistance and the kinetic friction to the classical mechanics homework-problem: finding the motion of a body sliding down a hemispherical hill. For a physically realistic ($\propto v^2$) form of drag force thus modified problem interestingly turns out to still admit analytical solutions of motion equations. I counter-intuitively find that even if friction coefficients are much larger than those of drag (as compared in the appropriate units), the drag effects remain significant. I also discuss how the transition between two possible outcomes of the motion (flying off and halting) is modified in the presence of the drag force coefficient with respect to the friction-only problem. I examine the behavior of the system using both the perturbation theory and exact analytical solutions of differential equations, showing that one can gain interesting insights from two different but complementary standpoints. 
\end{abstract}

\maketitle 

\input{body}

\appendix   

\input{appendices}

\begin{acknowledgments}
I would like to thank Dominik Kufel for all of his precious remarks and time dedicated to me as my scientific tutor. 
\newline
\newline
The author has no conflicts to disclose.
\end{acknowledgments}

\end{document}

%% file: body.tex
\section{Introduction - a model with friction and drag} 

A well-studied undergraduate homework problem in classical mechanics introduces a body (e.g. a skier) sliding down a frictionless spherical surface and asks where that body would fly off. It tests the student's ability to write down and scrutinize the equations of motion. More inquiring ones, however, could ask how close the solutions obtained are to the real situation of the skier on the hill. After all, any real skier knows, that friction and drag are quite crucial factors. It has been already proved that this problem's rough surface version still admits an analytical solution.\cite{Kufel,Mungan,Gonzales,Prior} In this article, I discuss one further generalization: the influence of adding air resistance to the problem, which for physically realistic ($\sim v^2$) form of drag force interestingly turns out to also admit analytical solution. However, I still assume that the hill is quarter-spherical (an interesting extension of that problem to a motion on an arbitrary curve with friction can be found in Ref. \onlinecite{Gonzales}).
\begin{figure} 
	\includegraphics[width= 0.55\columnwidth]{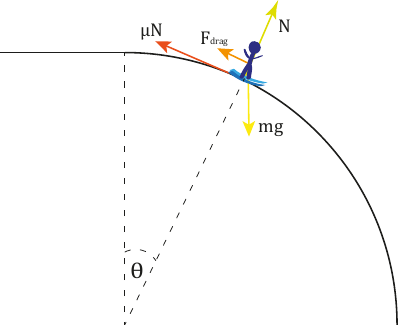}
	\caption{
		\justifying
		Forces acting on a skier on the hemispherical hill. In the radial direction: ground reaction force $N$; in the tangential direction: drag force $F_{\text{drag}}$ and friction force $\mu N$; vertically down: gravity force $m g$.}
	\label{Model}
\end{figure}
\\
We construct a model, visualized in the Fig. \ref{Model}. Assuming a turbulent flow of air, we can use a well-known expression for the drag force
$\vec{F}_{\text{drag}}=-\frac{1}{2}\rho v^2 C_{d} S_{d}\hat{v}\stackrel{def}{=}-v^{2}C\hat{v}$,
where $\rho$ is the air density, $C_{d}$ is the drag coefficient and $S_{d}$ is the cross-sectional area of the skier.
As long as the skier presses the hill (until they fly off), the hill acts on them with a normal reaction force $N$.
Including drag and kinetic friction ($\vec{F}_{\text{fr}} = - \mu N \hat{v}$) forces leads to the following radial and tangential components of the equations of motion:
\begin{eqnarray}
		N - m g \cos\theta = -m R \dot{\theta}^{2} \label{MotionEquations1} \text{;} \\
		m g \sin\theta - \mu N - C R^{2} \dot{\theta}^{2} = m R \ddot{\theta} \label{MotionEquations2}
\end{eqnarray}
(With boundary conditions $N>0$ and $\dot{\theta}\geq0$).\\
Differentiating the first equation with respect to time and dividing both sides by $\dot{\theta} \neq 0$ leads to
\begin{equation}
	\frac{dN}{d\theta} + m g\sin\theta = -2 m R \ddot{\theta}\text{.}
\end{equation}
It is quite easy to notice, that we can eliminate time derivatives of $\theta$ with appropriate substitutions from  Eq. (\ref{MotionEquations2}) and get a first-order linear equation for $N(\theta)$. This extends the method developed in Refs. \onlinecite{Kufel,Mungan, Prior} to the case with a drag force by observing that the drag force dependence term ($\sim v^2$) is analogous to the centripetal acceleration term (also $\sim v^2$) and thus still admits an analytical solution.  We define a dimensionless parameter  $A\stackrel{def}{=}\frac{\rho 	C_{d} S_{d} R}{m}=\frac{2 C R}{m}$ that characterizes the drag force in the system and get:
\begin{equation}\label{DiffEqua_N}
	\frac{dN}{d\theta} - N\left(2 \mu - A\right) = mg\left(A \cos\theta - 3 \sin\theta\right) \text{.}
\end{equation}
To obtain velocity as a function of angle, which is more crucial to our further analysis, it is convenient to rewrite Eqs. (\ref{MotionEquations1}) and (\ref{MotionEquations2}) according to the convention laid out in Ref. \onlinecite{Kufel}, thus introducing dimensionless quantities: $\Lambda\stackrel{def}{=}\frac{R \dot{\theta}^{2}}{g}$ and $\lambda\stackrel{def}{=}\frac{R \dot{\theta}_0^2}{g}$. Not only $\Lambda$ is a simple variable, characterizing the velocity of the skier, but also has a physical interpretation: it is a ratio of skier's instantaneous centripetal acceleration to the gravitational acceleration $g$.  Along that convention the first equation of motion (\ref{MotionEquations1}) takes the form: 
\begin{equation}\label{N}
	N = m g (\cos\theta - \Lambda) \text{.}
\end{equation}
We can differentiate that with respect to $\theta$ to get:
\begin{equation}\label{dN/dtheta}
	\frac{dN}{d\theta} = -m g \left( \sin\theta + \frac{d\Lambda}{d\theta}\right) \text{.}
\end{equation}
Inserting RHS of Eqs. (\ref{N}) and (\ref{dN/dtheta}) into Eq. (\ref{DiffEqua_N}) we obtain:
\begin{equation}\label{DiffEqua_Lambda}
	\frac{d\Lambda}{d\theta}-\Lambda(2 \mu - A) = 2\sin\theta - 2 \mu \cos\theta \text{.}
\end{equation}
Equation (\ref{DiffEqua_Lambda}) is accompanied with the boundary conditions:
\begin{eqnarray}
	\Lambda (0) &=& \lambda \text{,}\\
	0 ~\leq~ \Lambda(\theta) &<& \cos\theta \label{cond} \text{,}\\ 
	0 ~ \leq ~ \theta &<& \frac{\pi}{2} \text{.}
\end{eqnarray}
RHS of the inequality (\ref{cond}) comes from the condition $N > 0$ and Eq. (\ref{N}) -- our model holds only for a body that is still touching the hemisphere.\\
We notice that Eq. (\ref{DiffEqua_Lambda}) is a nonhomogeneous first-order ordinary differential equation. It can be solved fully analytically by the method of integrating factor.\cite{math} As a result of such calculations we get:
\begin{eqnarray}\label{LambdaExactSolution}
		\Lambda(\theta) &=& e^{\theta  (2 \mu -A)} \left( \frac{2A\mu -4 \mu^2+2}{(2 \mu -A)^2+1}+\lambda\right) \nonumber \\
		&& 
		 +\cos\theta \left( \frac{-2 A \mu +4 \mu ^2-2}{(2 \mu -A)^2+1}\right)    \nonumber \\
		 && + \sin\theta \left( \frac{2A-6\mu}{(2 \mu-A )^2+1}\right) \text{.}
\end{eqnarray}
As required, for $A=0$ this equation reduces itself to equations (12), (23), and  (A3) in Refs. \onlinecite{Kufel},\onlinecite{Prior}, and \onlinecite{Mungan}  respectively - solutions to this problem without drag force. Furthermore, nonzero drag $\Lambda(\theta)$ curves visible in Fig. \ref{fig1} in comparison to $A=0$ ones behave according to   intuition -  body affected by a nonzero drag tends to slow down more quickly.\\ \linebreak
It might have been argued, that for a large $\mu$ and a small $A$, correction for drag should not affect skier's motion significantly. Counterintuitively, it turns out to be not true in general (i.e. there exist initial velocity $\lambda \rightarrow\lambda_{\text{Amp}}$ for which even a small $A$ cannot be neglected). This effect is shown in the Fig. \ref{fig1}. To get further insight into this phenomenon I present the perturbative solution (with $A$ as the perturbation) to the Eq. (\ref{DiffEqua_Lambda}) in section \ref{Sect_PertSol}. I make a comparison of analytical and perturbative solutions in section \ref{Sec_Comparison}.
\\
In section \ref{Section_LiminalMotion} I discuss the phenomenon of the discontinuous ``phase'' (i.e. the final outcome of the motion - halting or flying off) transition effect, described earlier in Refs. \onlinecite{Kufel} and \onlinecite{Prior_comment}. It concerns a discontinuity in the skier's final angle as the function of an initial velocity (visible in Fig. \ref{fig1}). We will interestingly find that not only is the ``liminal $\Lambda(\theta)$'' (i.e. the one that separates motion's phases) less sensitive to the drag correction than any other $\Lambda(\theta)$, but also that $\Lambda(\theta)$ curves lying in the close neighborhood of the liminal one tend to be (surprisingly) the most affected ones by the nonzero $A$ inclusion.
\begin{figure}
	\includegraphics[width=\columnwidth]{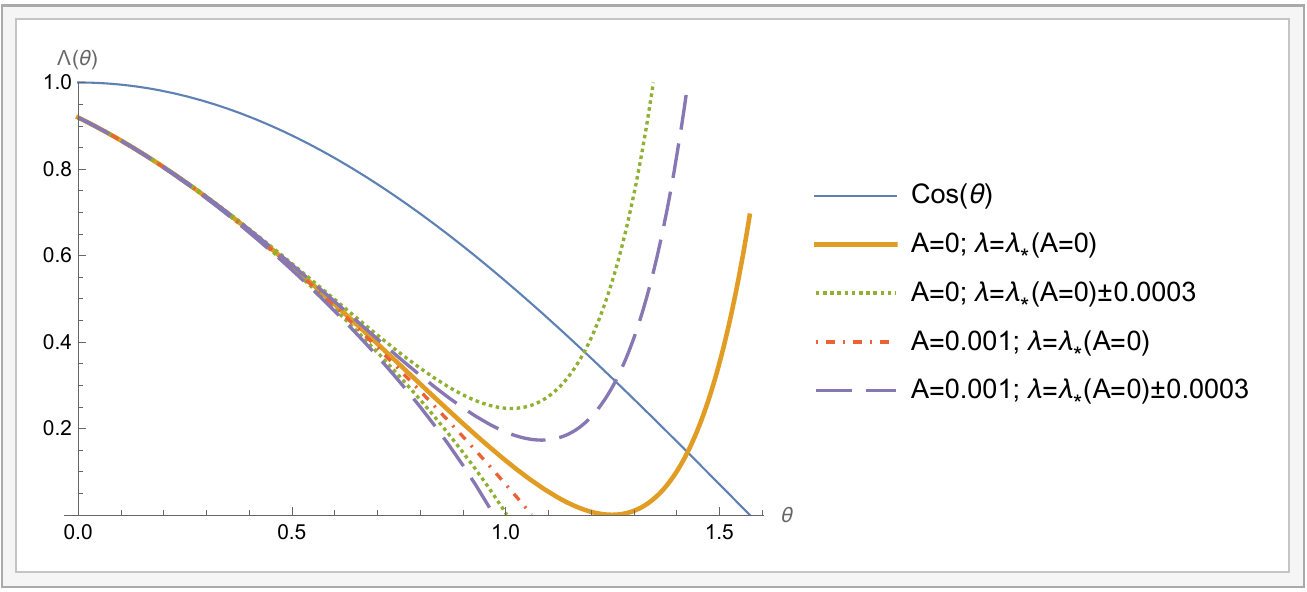}
	\caption{\justifying
		Plot of the velocity-determining function $\Lambda(\theta)$ over $\theta \in [0,\pi/2]$.Trajectories on this figure were generated for $\mu=3$ using exact formula (\ref{LambdaExactSolution}). Initial velocity for all of the generated curves was set to be equal to $\lambda_*(\mu=3, A=0)\simeq0.91901$ (a quantity introduced later in section \ref{Section_LiminalMotion}).\newline
		Comparison of dashed, purple and dotted, green curves surprisingly shows that including even very small, nonzero drag may lead to significant changes in motion. We can also see that neglecting the drag force can have a substantial effect on the final angle (point of intersection with the thin, blue curve, i.e. when $\Lambda(\theta)=\cos\theta$ or with the horizontal axis, i.e. when $\Lambda(\theta)=0$). \newline
		Moreover, the effect of the discontinuous phase transition is clearly visible - even infinitesimally small variations of the trajectory, from the curve that separates two different outcomes of the motion (halting or flying off; see section \ref{Section_LiminalMotion}), would change that outcome and, therefore, the final angle macroscopically.
	}
	\label{fig1}
\end{figure}
\section{Perturbative Solution} \label{Sect_PertSol}
Our main objective is to satisfy an inquiring student question: What if the model from the homework problem were more realistic? To address it, however, we should first wonder whether such a generalization is really necessary - in other words: When does the previous model fail? After all, even though the air resistance is undeniably there, in many models of everyday situations it is often rightfully omitted as rather small and unnecessary. In the problem considered in this article, we are particularly interested in the general conditions, for which motion in the corrected (drag-including) model differs much from one in the initial (drag-excluding) one, even if the drag is small ($A\ll1$). If we find those, then the doubt in the drag omission may be concerned as justified.\\
In order to investigate original model robustness against air resistance inclusion it is quite natural to use the perturbation theory. Number of terms in the $\Lambda(\theta)$ perturbative series in $A$, required to approximate the exact solution to Eq. (\ref{DiffEqua_Lambda}) well enough, will roughly tell us how sensitive to changes in $A$ the $\Lambda(\theta)$ curves are.
\subsection*{Parametrization with small $A$}
To estimate how adding drag force will affect the model with sole friction we expand $\Lambda(\theta, A, \mu)$ in the power series in $A$:
\begin{eqnarray}
	\Lambda(\theta, A, \mu) &=& \Lambda|_{A=0} + A 	\frac{\partial\Lambda}{\partial A}|_{A=0} + \frac{1}{2!}A^{2}\frac{\partial^{2}\Lambda}{\partial A^{2}}|_{A=0} + \dots \nonumber
	\\
	\label{LambdaExpansion}
	&=& b_{0} + A b_{1} + A^2 b_{2} + \dots~ \text{.}
\end{eqnarray}
We remember that by definition for any $A$ (in particular for $A=0$) and $\mu$:
\begin{equation}\label{LambdaInitialCond}
	\Lambda(\theta=0, A, \mu) =\lambda \text{.}
\end{equation}
Therefore we get the initial conditions:
\begin{eqnarray} 
	b_{0}(0)&=&\lambda \text{,} \nonumber\\
	b_{1}(0)&=&0\text{,} \nonumber\\
	b_{2}(0)&=&0\text{,} \label{b_InitCond} \\
	&\dots& \nonumber \\
	b_{n}(0)&=&0 \text{.} \nonumber
\end{eqnarray}
\\
Having inserted $\Lambda(\theta,A,\mu)$ from Eq. (\ref{LambdaExpansion}) into Eq. (\ref{DiffEqua_Lambda}) and extracting the terms of order $A^{0}$, we obtain a differential equation:
\begin{equation}\label{DiffEqua_b0}
	\frac{db_{0}}{d\theta} - 2\mu b_{0} = 2\sin\theta - 2\mu \cos\theta \text{.}
\end{equation}
This zeroth-order equation (here I call the equation order, the order of extracted terms from which it was obtained), accompanied  with the first boundary condition from eqns. (\ref{b_InitCond}), represents the situation for no drag force at all and its solution is just the $A=0$ case of the equation (\ref{LambdaExactSolution}):
\begin{eqnarray}\label{b0_theta}
			b_{0}&=&e^{2 \theta  \mu } \left(\lambda +\frac{3}{4 \mu ^2+1}-1\right)+\sin\theta\left( \frac{-6\mu}{4 \mu ^2+1}\right) \nonumber\\ 
			&& +\left(1-\frac{3}{4 \mu ^2+1}\right) \cos\theta \text{.}
\end{eqnarray}
Introducing constants:
\begin{eqnarray}
	\beta_{1}^0 &=& \frac{-6\mu}{4 \mu ^2+1} \text{;} \\
	\beta_{2}^0 &=& 1-\frac{3}{4 \mu ^2+1} \text{,}
\end{eqnarray}
we rewrite Eq. (\ref{b0_theta}) in the form:
\begin{equation} \label{b0_theta_InTermsOfBeta}
	b_{0}(\theta)=\left(\lambda -\beta_{2}^{0}\right) e^{2\theta\mu}+\beta_{1}^{0}\sin\theta+\beta_{2}^{0}\cos\theta \text{.}
\end{equation}
Superscript in $\beta_{i}^{0}$ denotes a degree of that constant (introduced later).\\
The right-hand side of the equation (\ref{DiffEqua_b0}) is specific for zeroth-order equation and for any higher order ($n>0$) we get:
\begin{equation}\label{DiffEqua_bn}
	\frac{db_{n}}{d\theta} - 2\mu b_{n} = - b_{n-1} \text{.}
\end{equation}
Knowing $b_{0}$, by iteration, we can obtain any further expansion term $b_{n}$. Since Eq. (\ref{DiffEqua_bn}) remains a linear differential equation, by the means of the method of integrating factor and having regarded appropriate initial condition (taken from Eqs. (\ref{b_InitCond})), we get the recurrence relation:
\begin{equation}\label{bn_integral}
	b_{n}(\theta)=-e^{2\mu\theta}\int_{0}^{\theta}e^{-2\mu t}b_{n-1}(t) \, dt \text{.}
\end{equation}
Now we derive the explicit formula for $b_{n}(\theta)$. First, we integrate $b_0(t)$ from the Eq. (\ref{b0_theta_InTermsOfBeta}) according to the Eq. (\ref{bn_integral}) (we remember that $\beta_1^0$ and $\beta_{2}^0$ are $t$-independent by definition):
\begin{eqnarray}\label{b1_theta}
	b_{1}(\theta)  &=& 
	\left( -\lambda\theta + \beta_{2}^{0}\theta -\frac{2\mu\beta_{2}^{0}+\beta_{1}^{0}}{1+4\mu^2}\right) e^{2\theta\mu}  \nonumber \\ 
	&& + \left(\frac{2\mu\beta_{1}^{0}-\beta_{2}^{0}}{1+4\mu^2} \right)\sin\theta + \left(\frac{2\mu\beta_{2}^{0}+\beta_{1}^{0}}{1+4\mu^2} \right)\cos\theta
	\nonumber \\ &=& 
	\left( -\lambda\theta  + \beta_{2}^{0}\theta -\beta_{2}^{1}\right) e^{2\theta\mu} + \beta_{1}^{1}\sin\theta + \beta_{2}^{1}\cos\theta \text{.} \qquad \quad
\end{eqnarray}
We observe that $b_{1}(\theta)$ has a similar form to $b_{0}(\theta)$ - the only difference is that the $\beta$ constants are now different functions of $\mu$ and we have added up one additional term. Analogically:
\begin{equation}
	b_{2}(\theta) = \left( \lambda\frac{\theta^{2}}{2} - \beta_{2}^{0}\frac{\theta^{2}}{2}+\beta_{2}^{1}\theta-\beta_{2}^{2} \right) e^{2\theta\mu} + \beta_{1}^{2}\sin\theta + \beta_{2}^{2}\cos\theta
\end{equation}
and so on. We notice that in general:
\begin{eqnarray}
	b_{n}(\theta)& =& \left(\lambda(-1)^n\frac{\theta^{n}}{n!}- \sum_{k=0}^{n}\beta_{2}^{n-k}(-1)^{k}\frac{\theta^{k}}{k!}\right) e^{2\theta\mu} \nonumber \\ 
	&& + \beta_{1}^{n}\sin\theta + \beta_{2}^{n}\cos\theta \label{bn_gen} 
\end{eqnarray}
where (see equation (\ref{b1_theta})), for $n>0$:
\begin{align}
	\beta_{1}^{n+1} &= \frac{1}{1+4\mu^2}\left(2\mu\beta_{1}^{n}-\beta_{2}^{n}\right) \label{beta1_rec} \text{;} \\
	\beta_{2}^{n+1} &= \frac{1}{1+4\mu^2}\left(2\mu\beta_{2}^{n}+\beta_{1}^{n}\right) \label{beta2_rec} \text{.}
\end{align}
This system of recurrent relations can be solved (see appendix \ref{app_solving_reccurence_relation}) to obtain  expressions:
\begin{eqnarray}
	\beta_{1}^{n} &=& \mathfrak{Re} \left(\left(\frac{1}{2 \mu - i} \right)^{n}\left(\beta_{1}^{0} + i \beta_{2}^{0} \right) \right) \nonumber \\
	&=& \mathfrak{Re} \left(2\left(\frac{1}{2 \mu - i} \right)^{n+1}\left(-1 + i \mu	 \right) \right) \text{,} \label{beta1_explicite} 
	\\
	\beta_{2}^{n} &=& \mathfrak{Re} \left(\left(\frac{1}{2 \mu - i}  \right)^{n}\left(\beta_{2}^{0} - i \beta_{1}^{0} \right) \right)\nonumber\\
	 &=& \mathfrak{Re} \left(2\left(\frac{1}{2 \mu - i} \right)^{n+1}\left(i +  \mu	 \right) \right) \label{beta2_explicite}
\end{eqnarray}
which, together with Eq.(\ref{bn_gen}), give us an explicit formula to calculate $b_{n}(\theta)$ of any order without the need to calculate integral  (\ref{bn_integral}) $n$ times. It is not surprise that such a formula exists, because, as it is shown above, Eq. (\ref{DiffEqua_Lambda}) - that perturbative solution we are seeking - has also an exact analytical solution. \\
Formula (\ref{bn_gen}) can be written more elegantly using complex numbers, whose real parts are given by eqns (\ref{beta1_explicite}) and (\ref{beta2_explicite}) - let's denote them by $\widetilde{\beta_1^n}$ and $\widetilde{\beta_2^n}$ respectively:
\begin{eqnarray}
	\widetilde{\beta_1^n} &=& 2\left(\frac{1}{2 \mu - i} \right)^{n+1}\left(-1 + i \mu	 \right) \text{;} \\
	\widetilde{\beta_2^n} &=& 2\left(\frac{1}{2 \mu - i} \right)^{n+1}\left(i +  \mu	 \right) \label{beta2_c} \text{.}
\end{eqnarray}
First we notice that$ \widetilde{\beta_1^n}=i \widetilde{\beta_2^n}$, meaning that both these numbers have same moduli and form a straight angle on a complex plane. Then we rewrite two last terms of (\ref{bn_gen}) and apply Euler's formula:
\begin{eqnarray}
	\beta_{1}^{n}\sin\theta + \beta_{2}^{n}\cos\theta &=& \mathfrak{Re}
	\left(\widetilde{\beta_1^{n}}\sin\theta +\widetilde{\beta_2^{n}}\cos\theta\right)\nonumber\\
	&=&\mathfrak{Re}
	\left(\widetilde{\beta_2^{n}}\left(\cos\theta+i\sin\theta\right)\right)\nonumber \\
	&=&\mathfrak{Re}\left(\widetilde{\beta_2^{n}}\, e^{i\theta}\right) \text{.}
\end{eqnarray}
Next we perform some manipulations on the part of the exponential term of formula (\ref{bn_gen}):
\begin{eqnarray}
	\sum_{k=0}^{n}\beta_{2}^{n-k}(-1)^{k}\frac{\theta^{k}}{k!} &=& \mathfrak{Re}\left( \sum_{k=0}^{n}\widetilde{\beta_2^{n-k}}(-1)^{k}\frac{\theta^{k}}{k!}\right)\nonumber\\
	&=&\mathfrak{Re}\left(\widetilde{\beta_2^{n}}\sum_{k=0}^{n}\frac{(i\theta -2\mu\theta)^k}{k!}\right)\text{.} \qquad
\end{eqnarray}
And finally rewrite (\ref{bn_gen}) in the form of:
\begin{eqnarray}
	b_n(\theta) &=& \mathfrak{Re}\left(
	\lambda\frac{(-\theta)^n}{n!} +
	\widetilde{\beta_2^{n}}\left(e^{(i\theta-2\mu\theta)}  \vphantom{\int_{min}^{max}} \right. \right. \nonumber \\
	&&\left. \qquad \qquad \qquad \quad- \left.\sum_{k=0}^{n}\frac{(i\theta   -2\mu\theta)^k}{k!}\right)
	\right)e^{2\mu\theta} \label{bn_gen_c_alt} \\
	&=& \mathfrak{Re}\left(
	\lambda\frac{(-\theta)^n}{n!} +
	\widetilde{\beta_2^{n}}\sum_{k=n+1}^{\infty}\frac{(i\theta -2\mu\theta)^k}{k!}\right)
	e^{2\mu\theta} \label{bn_gen_c} \text{.} \qquad
\end{eqnarray}
\section{Comparison of exact and perturbative solutions} \label{Sec_Comparison}
First and foremost, it should be underlined, that the perturbative solution converges to the exact one. Even though the subsequent terms in this expansion may look quite complicated, we can represent the whole perturbative series as the sum of the simpler subseries, that turn out to reduce to the geometric series sum or exponential function expansion formulae. Careful simplifications will lead to Eq. (\ref{LambdaExactSolution}) (see appendix \ref{app_pertu_to_exact}).
\subsection{Correctness of the approximation}
It is important to answer the question, how good approximations given by the perturbative approach are -- that is how many perturbative terms do we need to solve the problem within a certain degree of accuracy. To compare the magnitudes of succeeding perturbative terms, we need to estimate values of expressions $b_{n}/b_{n-1}$. From expression (\ref{bn_gen_c}) we have:
\begin{widetext}
	\begin{eqnarray}
		b_{n+1}(\theta) &=& \mathfrak{Re}\left(
		\lambda\frac{(-\theta)^{n+1}}{(n+1)!} +
		\widetilde{\beta_2^{n+1}}\sum_{k=n+2}^{\infty}\frac{(i\theta -2\mu\theta)^k}{k!}\right)
		e^{2\mu\theta}  \nonumber\\
		&=&\mathfrak{Re}\left(
		\lambda\frac{(-\theta)^n}{n!}\frac{(-\theta)}{n+1}  +
		\frac{1}{2\mu - i}\widetilde{\beta_2^{n}} \sum_{k=n+1}^{\infty}\frac{(i\theta -2\mu\theta)^{k}}{k!} \frac{(i\theta-2\mu\theta)}{k+1}\right)
		e^{2\mu\theta}  \nonumber\\
		&=& \frac{(-\theta)}{n+1}
		\mathfrak{Re}\left(
		\lambda\frac{(-\theta)^n}{n!}  +
		\widetilde{\beta_2^{n}} \sum_{k=n+1}^{\infty}\frac{(i\theta -2\mu\theta)^{k}}{k!} \frac{(n+1)}{k+1}\right)
		e^{2\mu\theta}  \nonumber\\
		&<& \frac{(-\theta)}{n+1} b_n(\theta) \text{.}
	\end{eqnarray}
\end{widetext}
So the ratio of the subsequent terms of the expansion given by Eq. (\ref{LambdaExpansion}) satisfies the inequality:
\begin{equation}
	\left|\frac{A^{n+1}b_{n+1}(\theta)}{A^{n}b_{n}(\theta)}\right| < \frac{A\theta}{n+1} \text{.} \label{bn_Ration_MiuIntermediate} 
\end{equation}
We observe that our approximation worsens with angle, but works well for $A < (n+1)/\theta$.\\ 
For a large $\mu$, as expected, approximation gets better. As $\mu \rightarrow \infty$, $b_{n} \rightarrow e^{2\theta\mu}(\lambda-1)((-\theta)^{n}/n!)$ for $n>0$, so in this scenario:
\begin{equation}
	\left|\frac{A^{n+1}b_{n+1}(\theta)}{A^{n}b_{n}(\theta)}\right| = \frac{A \theta}{n+1} \text{.} \label{bn_Ration_MiuInf}
\end{equation}
Although this is a reasonable outcome - for a large value of friction we fairly do not need to worry about any drag force - in a moment we will see, that this outcome not always holds. \\
\subsection{Amplification of the first order perturbations} \label{Sec_Amplification}
\begin{figure} 
	\begin{subfigure}{0.49\textwidth}
		\includegraphics[width=\textwidth]{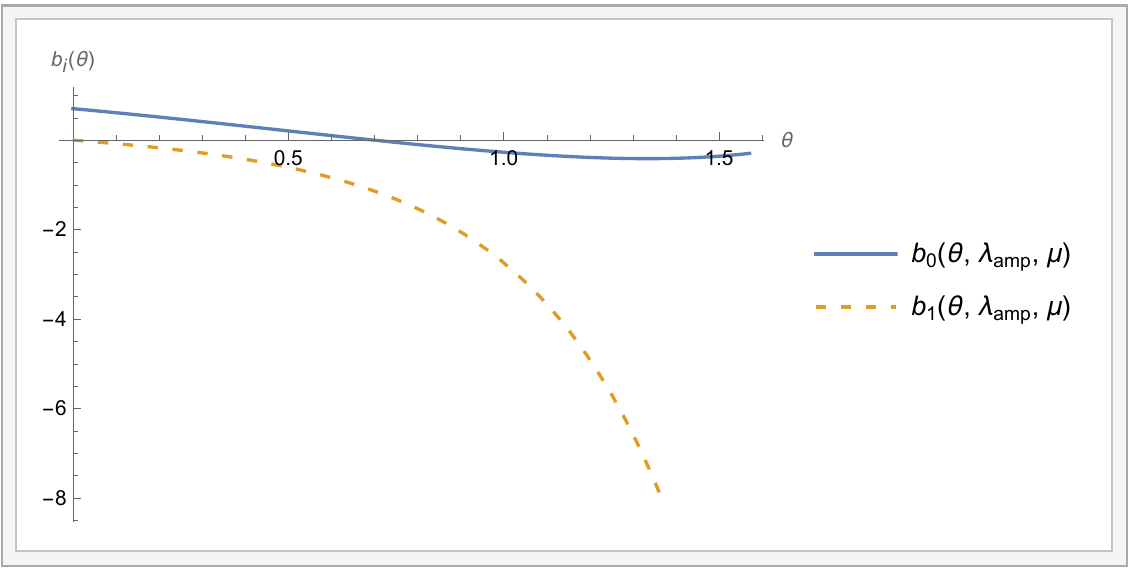}
		\caption{$\mu=1.5$}
	\end{subfigure}
	\begin{subfigure}{0.49\textwidth}
		\includegraphics[width=\textwidth]{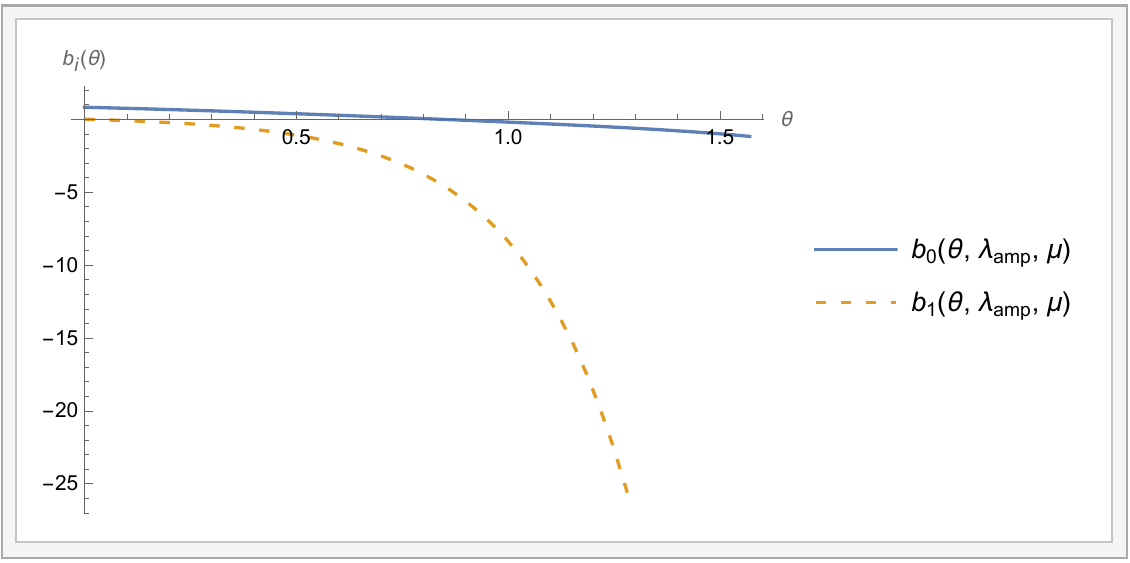}
		\caption{$\mu=2$}
	\end{subfigure}
	\begin{subfigure}{0.49\textwidth}
		\includegraphics[width=\textwidth]{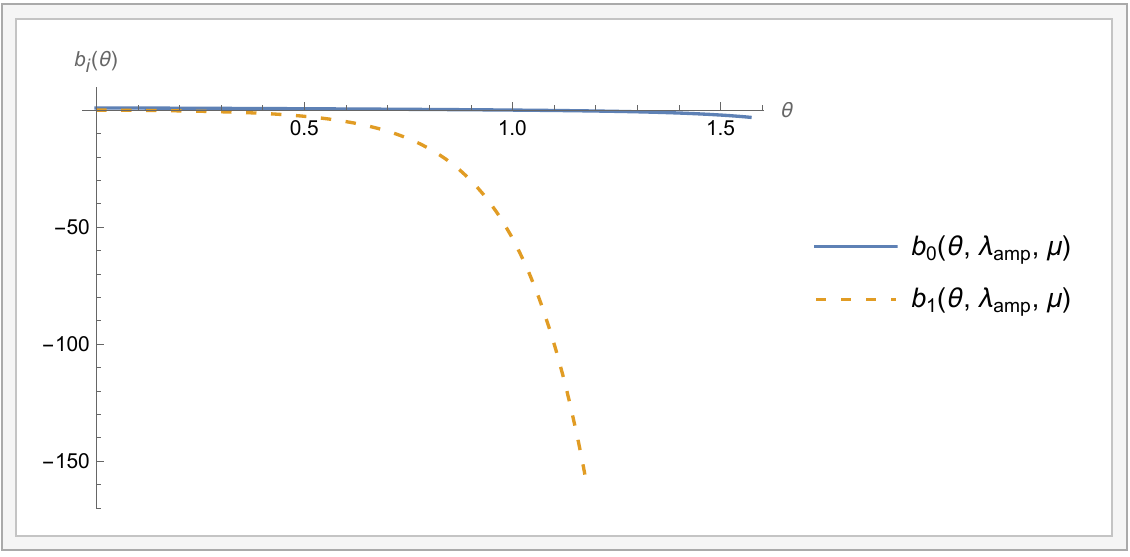}
		\caption{$\mu=3$}
	\end{subfigure}
	\caption{\justifying
		Comparison of zeroth ($b_0(\theta)$) and first ($b_1(\theta)$) perturbative terms for $\lambda=\lambda_{\text{Amp}}$. We see that the difference in their respective values grows with $\mu$.}
	\label{zeroth and first pert term for lamba=lambda_amp compared}
\end{figure}

\begin{figure}
	\begin{subfigure}{0.49\textwidth}
		\includegraphics[width=\textwidth]{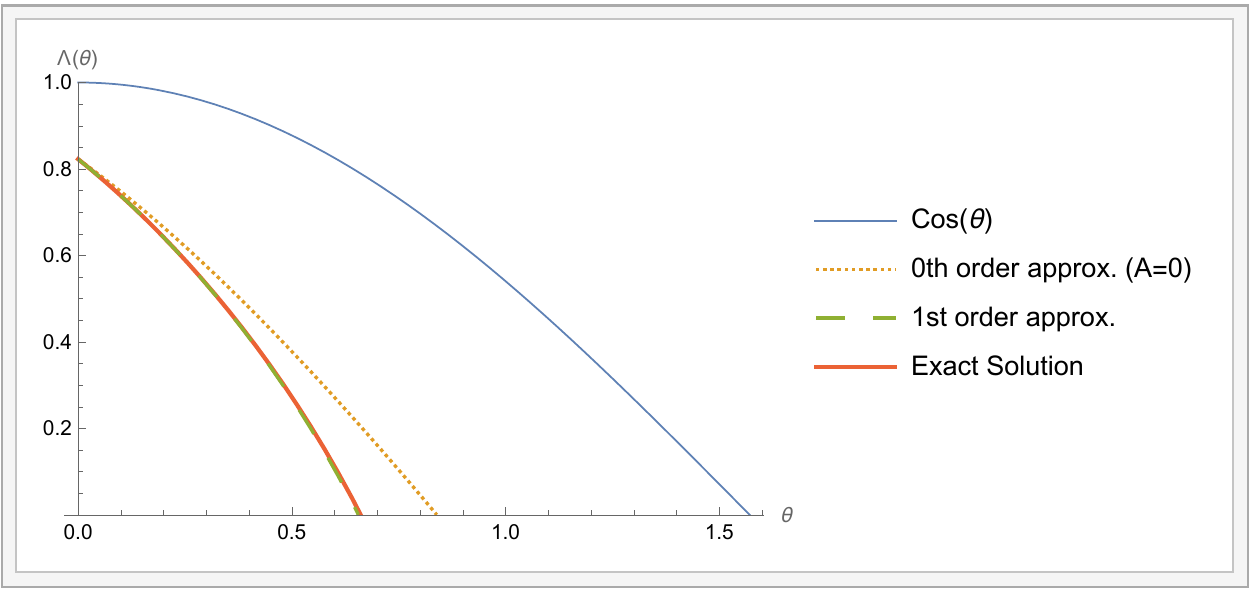}
		\caption{$\mu=2$ ~ $A=0.1$}
	\end{subfigure}
	\begin{subfigure}{0.49\textwidth}
		\includegraphics[width=\textwidth]{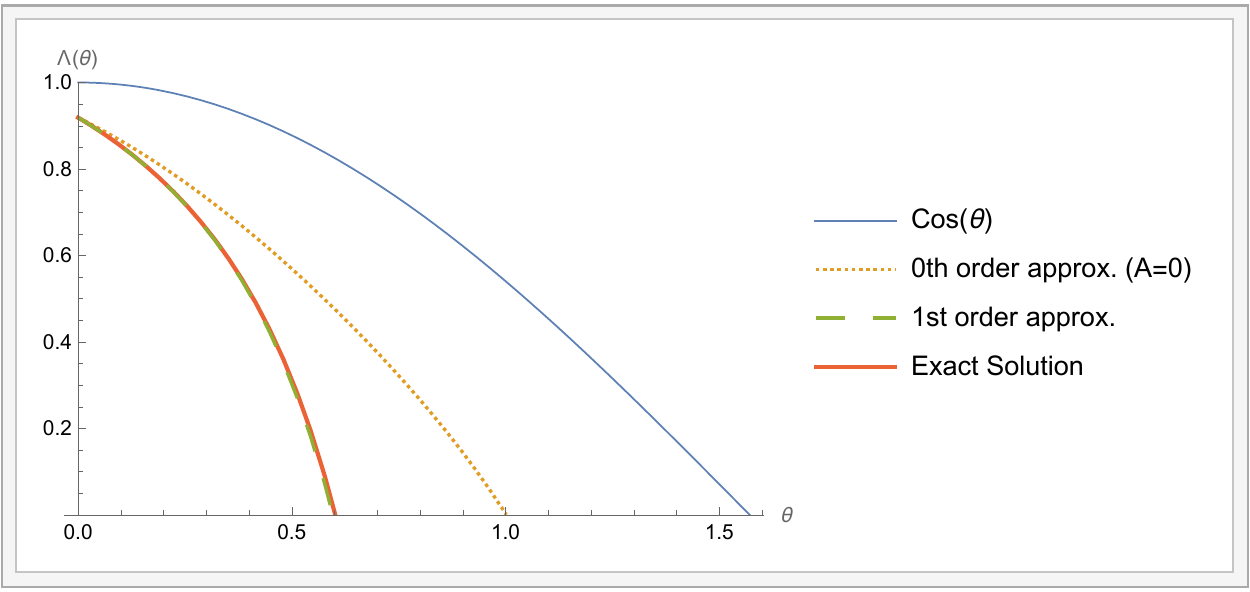}
		\caption{$\mu=3$ ~ $A=0.1$}
	\end{subfigure}
	\begin{subfigure}{0.49\textwidth}
		\includegraphics[width=\textwidth]{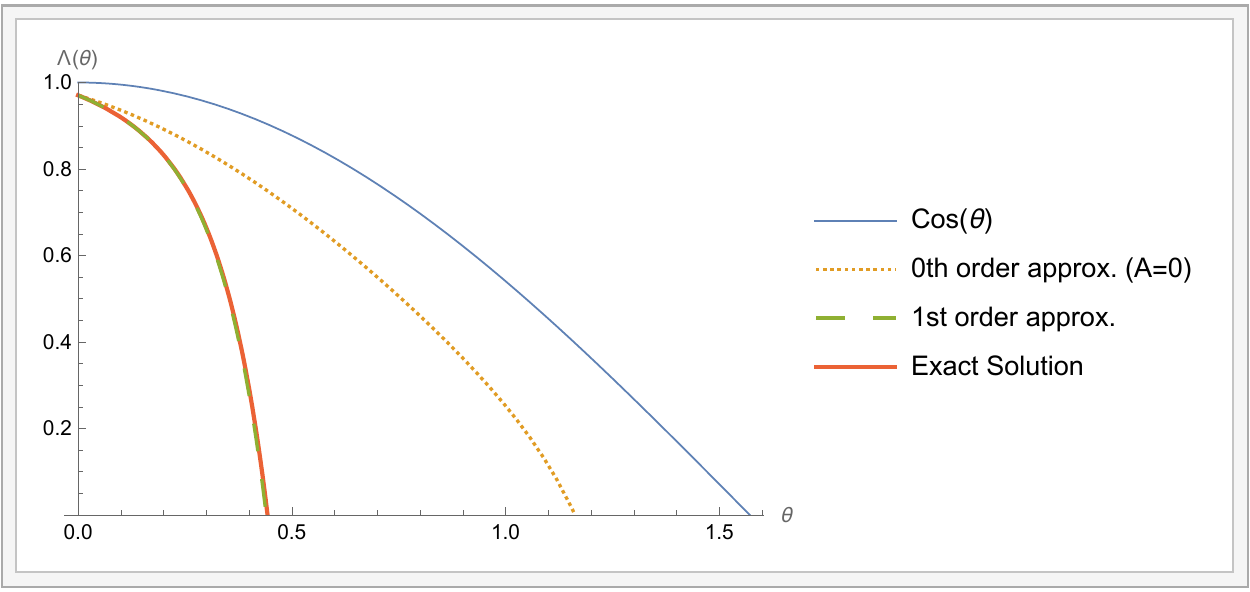}
		\caption{$\mu=5$ ~ $A=0.1$}
	\end{subfigure}
	\begin{subfigure}{0.49\textwidth}
		\includegraphics[width=\textwidth]{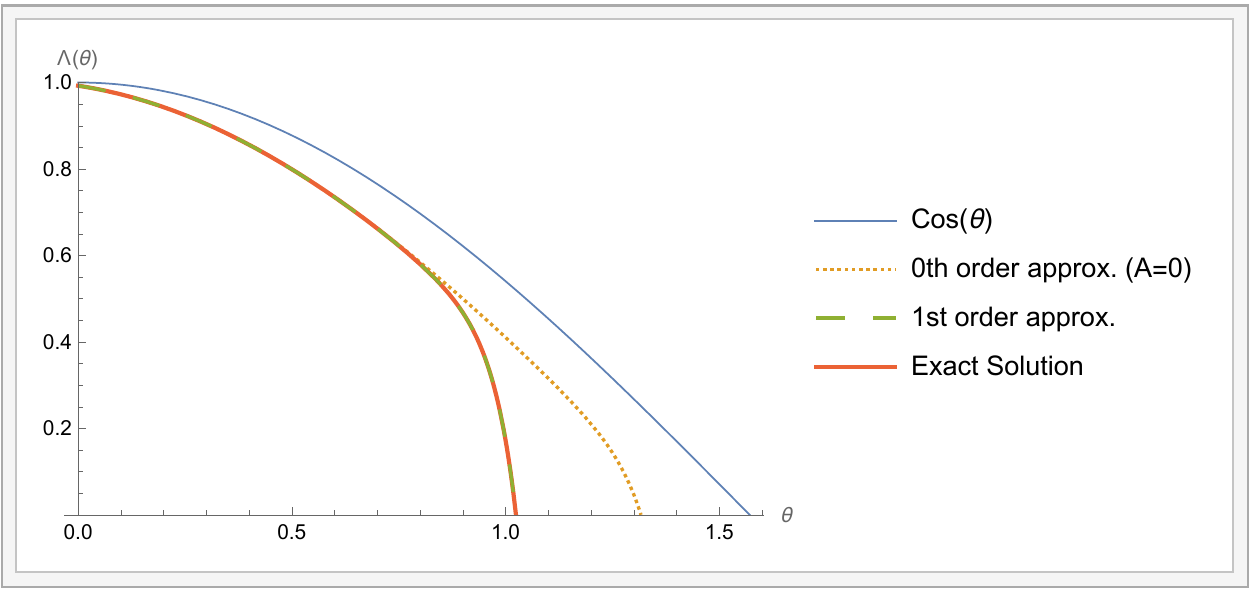}
		\caption{$\mu=10$ ~ $A=10^{-8}$}
		\label{1st_order_pert_d}
	\end{subfigure}
	\caption{\justifying
		Effects of amplification of first-order perturbations. 0-th order approximation curve corresponds to $A=0$ case, while 1-st order approx. curve is plotted having taken under consideration the approximation of order $n=1$ (along Eqs. (\ref{LambdaExactSolution})/(\ref{bn_gen})) for value of $A$ denoted under the plot. All plots were made for $\lambda=\lambda_{\text{Amp}}$.  \newline
	It can be seen that the magnitude of the effect grows with $\mu$ and $\theta$. On all plots dashed, green curves are almost collinear with thick, red ones.}
	\label{1st_order_pert}
\end{figure}

Let us wonder, under what conditions the system will be most prone to drag force inclusion. By that we mean finding the initial parameters ($\lambda,\mu$) for which $\Lambda(\theta)$ will change the most while adding drag force to the analysis. I conjecture, that it will be the case if the mean value of the function $b_0(\theta)$ on the  angles interval $[0,\pi/2]$ zeroes itself, hence making $b_1(\theta)$ term more significant:
\begin{equation}
	\int_{0}^{\pi/2} b_0(\theta)\ d\theta  = 0 \text{.} \label{1st_amplif_cond}
\end{equation}
Zeroth order term in the perturbative series takes form:
\begin{equation}
	b_0(\theta) = (\lambda - \beta_2^0)e^{2\mu\theta} + \mathfrak{Re}\left( \widetilde{\beta_2^{0}}e^{i\theta}\right) \label{b0_c}\text{.}
\end{equation}
Having substituted (\ref{b0_c}) into (\ref{1st_amplif_cond}) we solve for $\lambda$ and obtain, that (\ref{1st_amplif_cond}) is satisfied when:
\begin{eqnarray}\label{BigFirstPerturb_Cond}
	\lambda \rightarrow \lambda_{\text{Amp}} &=& \beta_{2}^{0} - \frac{2\mu (\beta_{1}^0 + \beta_2^0)}{e^{\pi\mu}-1} \nonumber \\ 
	&=& \frac{4 \mu ^2-2}{4 \mu ^2+1}-\frac{4 \mu  \left(2 \mu ^2-3 \mu -1\right)}{\left(e^{\pi  \mu }-1\right) \left(4 \mu ^2+1\right)} \text{.}
\end{eqnarray}
When condition (\ref{BigFirstPerturb_Cond}) holds true, the first-order term generally is much larger than the zeroth-order term (See Fig. \ref{zeroth and first pert term for lamba=lambda_amp compared}) and therefore the system is especially prone to drag-force perturbation (See Fig. \ref{1st_order_pert}).
\\
For large $\mu$, the RHS of the condition (\ref{BigFirstPerturb_Cond}) reduces to $\beta_{2}^0$ and the amplification effect may be interpreted as the result of cancellation of the exponent term in (\ref{b0_c}) - the exponential part does not vanish for the subsequent terms of perturbative series ($b_i(\theta)$) and since it is always the largest part of each term ($|\widetilde{\beta_2^n}|<2$), the ones succeeding $b_0(\theta)$ tend to be larger than the zeroth-order one. The magnitude of this effect is fairly proportional to the $e^{2\theta\mu}$. Therefore, what is somewhat surprising, the larger the $\mu$, the greater the first amplification effect (the difference between the first term and subsequent ones becomes more significant). This means, that for large frictions, we can find such an initial velocity $\lambda$, for which even a tiny drag force cannot be omitted, as it shall have a vast impact on the subsequent perturbative terms and therefore the skier's motion. Fig. \ref{1st_order_pert} shows visible effects of adding drag force, with drag parameter $A$ even as minor as $A =10^{-8}$ (subfigure \ref{1st_order_pert_d}). This presents an exception from the observation made after comparing expressions (\ref{bn_Ration_MiuIntermediate}) and (\ref{bn_Ration_MiuInf}) and constitutes an interesting example in which common intuition fails to work.\\
It also should be noticed, that $\lambda$ is defined, only in the interval $[0,1]$, so the effect of the perturbation amplification cannot occur for $\mu \lesssim 0.479$ (which was evaluated numerically having demanded that the RHS of expression (\ref{BigFirstPerturb_Cond}) is $> 0$).\\
This effect increases along $\theta$, so it is more noticeable for greater angles. As we will see in section \ref{Section_LiminalMotion} (and what is proved in appendix \ref{app_behavior_of_Lambda}), the skier achieves the largest angles when they start their motion with a velocity called liminal (and denoted by $\lambda_*$). It turns out to be true, that the amplification effect is most dramatic for motions near the liminal one (but still satisfying $\lambda \rightarrow \lambda_{\text{Amp}}$, so we conclude it is so when $\lambda_* \approx \lambda_{\text{Amp}}$), however, as it is later showed, it ceases for the liminal motion itself, as for that particular motion $e^{2\theta\mu}$ term disappears for every $b_n$. \\

\section{Phase-boundary effect and the liminal velocity} \label{Section_LiminalMotion}
We can now analyze the motion of our system in an $(A, \mu, \lambda)$ variable space. The skier starts with a velocity $\lambda$ and on one hand is being accelerated by gravity and on the other is being slowed down by friction and drag force. So $\Lambda(\theta)$ is bent up by the first factor and down by the second one. Depending on which of these two factors is more significant, the motion can end in one of two ways:
\begin{enumerate}
	\item The skier comes to rest when $\Lambda(\theta)=0$;
	\item The skier flies off the hill when $\Lambda(\theta)=\cos\theta$ (From condition $N=0$ and Eq. (\ref{N})).
\end{enumerate}
The liminal trajectory (phase-boundary curve) for a given set of $(A, \mu)$ occurs when the skier almost halts, but in the last moment his $\Lambda(\theta)$ function is bent upwards. Mathematically it means that for some angle $\theta_*$ both $\Lambda(\theta_*)=0$ and $\frac{d\Lambda}{d\theta}=0$ are valid. From this two conditions and Eq. (\ref{DiffEqua_Lambda}) we obtain, that:
\begin{equation}\label{theta*}
	\theta_*=\arctan\mu \text{,}
\end{equation}
what somewhat surprisingly agrees with the result found in Refs. \onlinecite{Kufel, Gonzales, Prior_comment} for a problem without drag. This means, that the appearance of drag force cannot influence a maximum halt angle (which happens to be equal to $\theta_{*}$, as it is shown in appendix \ref{app_behavior_of_Lambda}) achievable with fixed $\mu$. However it still both can and will change the initial velocity, required to achieve that angle. \\

\subsection{Perturbative approach}
\begin{figure}
	\begin{subfigure}{0.49\textwidth}
		\includegraphics[width=\textwidth]{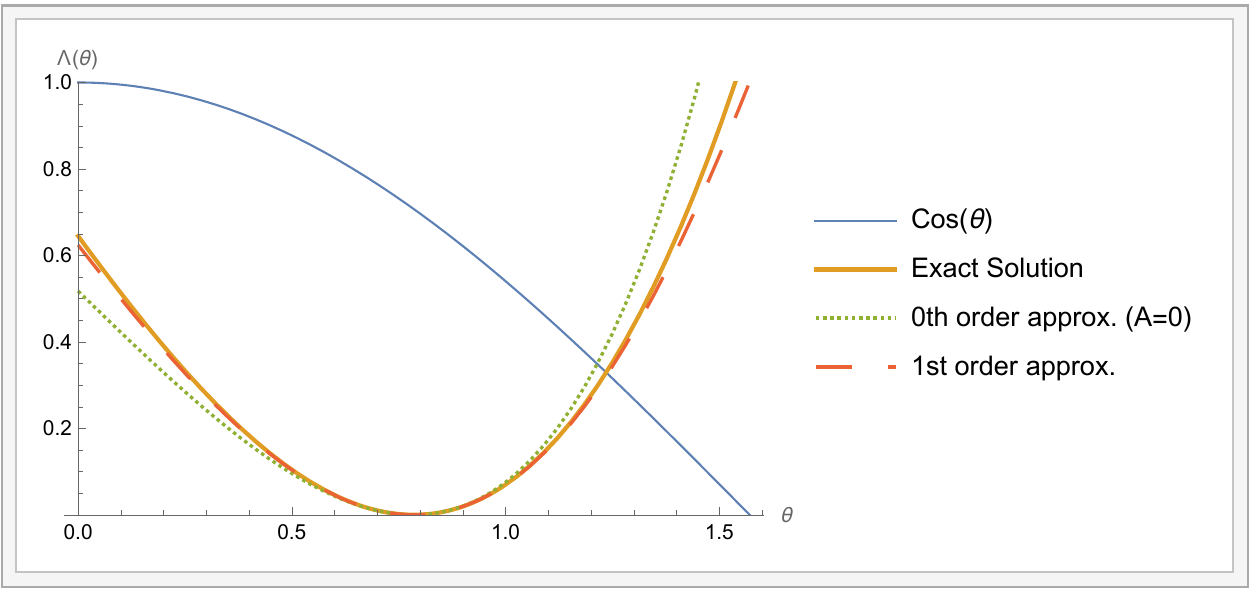}
		\caption{$\mu=1$ ~ $A=1$}
	\end{subfigure}
	\begin{subfigure}{0.49\textwidth}
		\includegraphics[width=\textwidth]{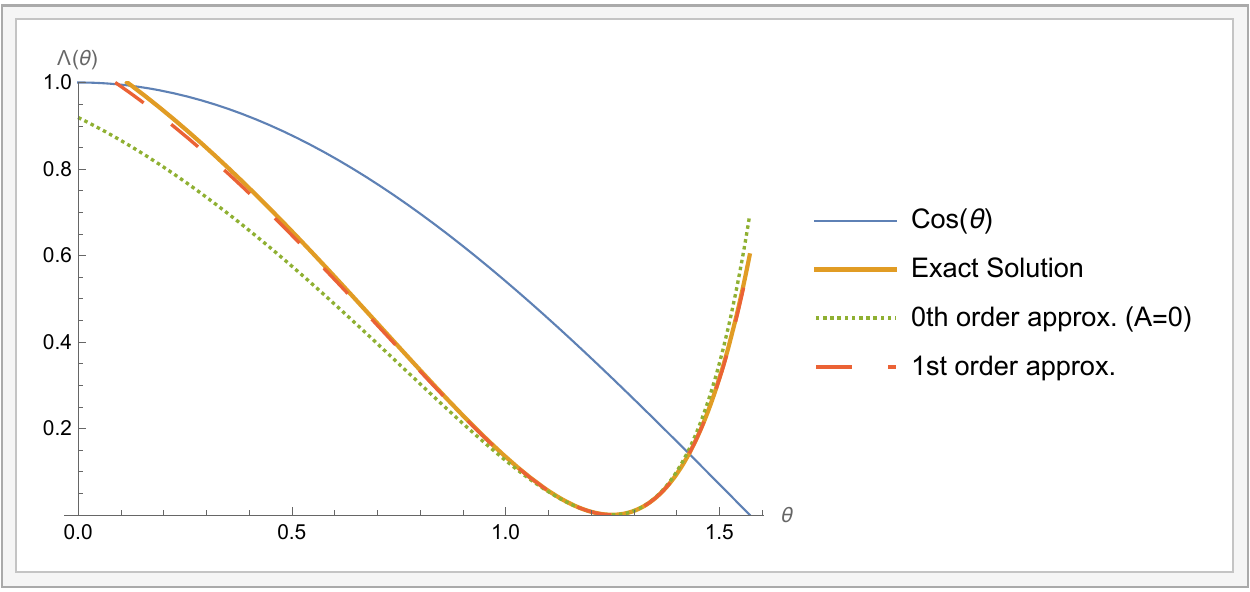}
		\caption{$\mu=3$ ~ $A=1$} \label{fig_liminal_motion_b}
	\end{subfigure}
	\caption{\justifying
		Approximations of liminal motion curve given by Eq. (\ref{Lambda_theta_lambda*PertExp}) together with exact solution $\Lambda(\theta,\lambda_{*})$ obtained with inserting exact value $\lambda_*$ (see Eq. (\ref{lambda_*_exact}) in section \ref{Sect_ExactSol}) into Eq. (\ref{LambdaExactSolution}). As we see the first-order approximation is nearly collinear with the exact solution curve and even zeroth-order approximation holds quite well for fairly significant $A=1$.\newline
	We can also observe, that in the figure \ref{fig_liminal_motion_b}, the liminal curve for nonzero $A=1$, represents the body, beginning its motion with $\lambda>1$, which is impossible, hence such a body would immediately fly off. It is a manifestation of fly-off phase disappearance effect for motion characterized with parameters $A=1$ and $\mu=3$.} \label{fig_liminal_motion}
\end{figure}

\paragraph{Finding the liminal motion}
Now we want to find an initial velocity $\lambda_*(A,\mu)$ such that, the skier starting theirs motion with it will be on the liminal trajectory. I will call it liminal initial velocity. Obtained from a perturbative expansion, it can be rightfully  anticipated to be in the form of a certain perturbative series (however, as we will see, this perturbative series will be equivalent to an exact solution for $\lambda_*$, that still exists)\\
To obtain an expression for $\lambda_*$, we substitute $\theta=\theta_*$ given by  Eq. (\ref{theta*}) into Eq. (\ref{LambdaExpansion}), accompanied by Eq. (\ref{bn_gen_c_alt}) and then compare the LHS to zero. From Eq. (\ref{bn_gen_c_alt}) we see, that for each $n$ there is only one occurrence of $\lambda_*$ - namely $\lambda_* e^{2\theta\mu}(-A\theta_*)^{n}/n!$. This expression summed up for $n$ from $0$ to infinity is nothing else but Taylor expansion over $A\theta$ for $\lambda_* e^{\theta_*(2\mu-A)}$. Having solved for $\lambda_*$:
\begin{widetext}
	\begin{eqnarray}
		\lambda_* &=& \mathfrak{Re}\left( -e^{\theta_*(A-2\mu)}\left(\sum_{n=0}^{\infty}A^n \widetilde{\beta_2^{n}} \left(e^{i\theta_*}-e^{2\mu\theta_{*}}\sum_{k=0}^{n}\frac{((i-2\mu)\theta_*)^k}{k!} \right)  \right)\right) \nonumber\\
		&=&\mathfrak{Re}\left(  -e^{-2\theta_*\mu}\left(\sum_{n=0}^{\infty}\frac{(A\theta_*)^n}{n!}\right)\left(\sum_{n=0}^{\infty}A^n \widetilde{\beta_2^{n}} \left(e^{i\theta_*}-e^{2\mu\theta_{*}}\sum_{k=0}^{n}\frac{((i-2\mu)\theta_*)^k}{k!} \right)  \right)\right)  \nonumber \\
		&=&\mathfrak{Re}\left( 
		\sum_{n=0}^{\infty}A^n \left(  \sum_{l=0}^{n} \frac{\theta_*^{n-l}}{(n-l)!} \widetilde{\beta_2^l} \left(\left(\sum_{k=0}^{l} \frac{((i-2\mu)\theta_*)^k}{k!}\right) - e^{(i-2\mu)\theta_*} \right)  \right)\right)   \nonumber\\
		&=&\mathfrak{Re}\left(
		\sum_{n=0}^{\infty}A^n \left( \widetilde{\beta_2^n} - e^{-2\mu\theta_{*}} \sum_{k=0}^{n} (\widetilde{\beta_2^{k}}e^{i\theta_{*}}) \frac{\theta_*^{n-k}}{(n-k)!}  \right) 
		\label{lambda*} \right) \\
		&=&\sum_{n=0}^{\infty}A^n \left( \beta_2^n	 - e^{-2\mu\arctan\mu} \sum_{k=0}^{n} \left(\frac{\beta_{1}^k \mu + \beta_{2}^k}{\sqrt{1+\mu^2}} \right)  \frac{(\arctan\mu)^{n-k}}{(n-k)!}  \right) \label{lambda*_pert}
	\end{eqnarray}
\end{widetext}
(penultimate equality can be proved by induction).
\\
Now, to say something about the influence of adding drag force on the liminal motion itself, we calculate $\Lambda(\theta, \lambda_*)$. Inserting  $\lambda=\lambda_{*}$ given by Eq. (\ref{lambda*}) into the original expansion (\ref{LambdaExpansion}) and performing some rearrangements leaves us with:
\begin{widetext}
	\begin{eqnarray} \label{Lambda_theta_lambda*PertExp}
		\Lambda(\theta, \lambda_*) &=& \mathfrak{Re}\left( 
		\sum_{n=0}^{\infty}A^{n} \left( -\left( \sum_{k=0}^{n} \widetilde{\beta_2^{k}}e^{i\theta_{*}}  \frac{(\theta^*-\theta)^{n-k}}{(n-k)!} \right)e^{2\mu(\theta-\theta_{*})} + \widetilde{\beta_2^{n}}e^{i\theta} \right)\right) \nonumber \\
		&=& 
		\sum_{n=0}^{\infty}A^{n} \left( -\left( \sum_{k=0}^{n} \frac{\beta_{1}^k \mu + \beta_{2}^k}{\sqrt{1+\mu^2}} \frac{(\arctan\mu-\theta)^{n-k}}{(n-k)!}  \right)e^{2\mu(\theta-\arctan\mu)} + \beta_{1}^n \sin\theta + \beta_{2}^n  \cos\theta  \vphantom{\sum_{min}^{max}}\right) \text{.}
	\end{eqnarray}
\end{widetext}
This solution allows us to draw some solid conclusions. Firstly, if $\theta \rightarrow \theta_*$, all terms of the power series (\ref{Lambda_theta_lambda*PertExp}) converge to zero. For $\theta$ far away from $\theta_{*}$, the largest term happens to be the one with exponent, and using similar reasoning as in the previous section, we reach the conclusion:
\begin{equation}
	\left|\frac{A b_{n+1}(\lambda_*,\theta)}{b_{n}(\lambda_*,\theta)}\right| < \frac{A (\theta-\arctan\mu)}{n+1} \text{.}
\end{equation}
This outcome, compared with the Eq. (\ref{bn_Ration_MiuIntermediate}), says that the liminal curve is always less prone to an $A$ influence than any chosen non-liminal curve.\\
Furthermore, we can observe that for $\Lambda(\theta, \lambda_*)$ the first order amplification effect, described earlier, takes place only for $\mu \simeq 2$ (calculated numerically from comparing mean value of the $A^0$ perturbative term of expansion ($\ref{Lambda_theta_lambda*PertExp}$) to zero) and cannot occur for large $\mu$ as then its $e^{2\mu\theta}$ term vanishes for all $n$. All of these effects are visible in Fig. \ref{fig_liminal_motion}.\\
Even though, for large $\mu$, there can be no first-order amplification for $\Lambda(\theta, \lambda_*)$, we know that it does happen for $\Lambda(\theta, \lambda=\beta_{2}^0 )$. As it has been discussed in the previous section, if this amplification occurs for an almost liminal curve ($\lambda=\beta_{2}^0 \text{ near } \lambda_*$), it shall be more visible, as such a motion would allow greater angles (see appendix \ref{app_behavior_of_Lambda}). Let us regard, under what conditions $\lambda_*\sim\beta_{2}^0$. From the Eq. (\ref{lambda*_pert}) it can be deduced, that this happens if simultaneously $\mu$ increases and $A$ declines. Therefore we conclude, that the influence of neglecting $A$ should be most visible if the friction was huge and the disregarded drag wasn't too large. This finding even intensifies the counter-intuitive upshot of the previous section and justifies the tendency that may be observed in the Fig. \ref{1st_order_pert}.

\paragraph{Fly-off phase forbiddance}
\begin{figure}
	\includegraphics[width=\columnwidth]{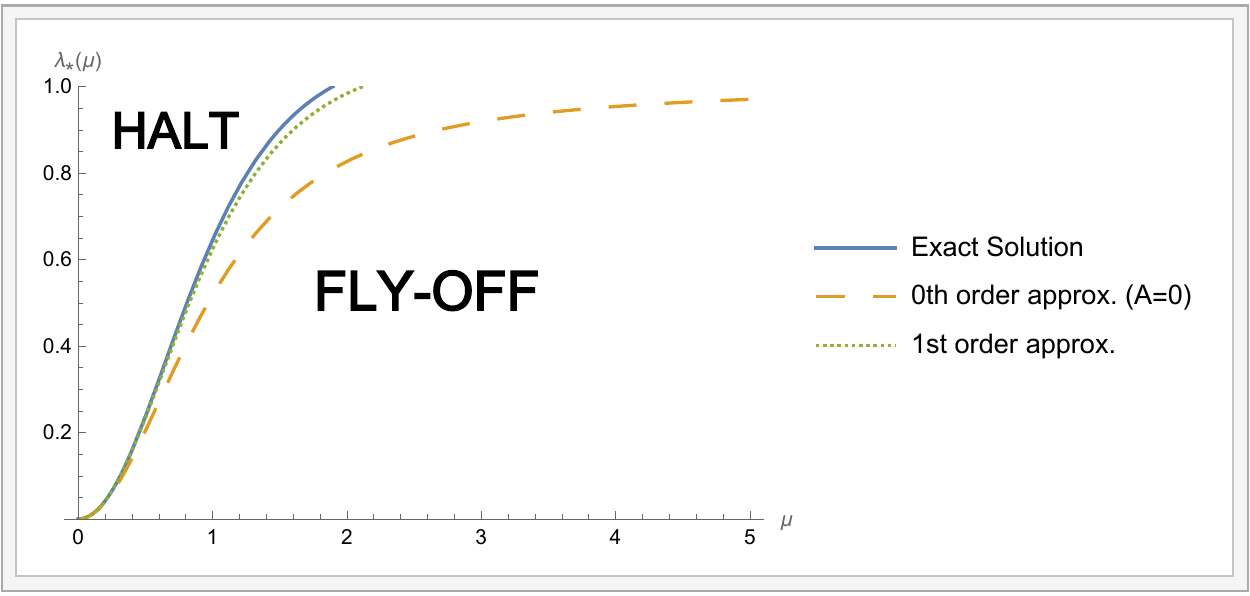}
	\caption{\justifying
		The curve $\lambda_{*}(\mu)$, that forms the boundary between the "halt" and "fly-off" phases. It was generated for $A=1$, using exact solution (\ref{lambda_*_exact}), and perturbative approximation (\ref{lambda*_pert}) taken up to the zeroth and first order terms.}
	\label{fig_phase_div}
\end{figure}
It can be observed (see figures \ref{fig_phase_div} and \ref{fig_liminal_motion_b}), that for sufficiently high $\mu$ and $A$, $\lambda_{*}$ given by Eqn. (\ref{lambda*_pert}) can exceed range $1$. For large $\mu$, the perturbative expansion up to the first term approximately takes form $\lambda_{*} \simeq \beta_{2}^0+ A \beta_{2}^1$ (the remaining parts converge to 0 exponentially). Solving inequality $\beta_{2}^0+ A \beta_{2}^1>1$ gives us (assuming large $\mu>\sqrt{5}/2$): $A>(3+ 12 \mu^2)/(-10\mu+8\mu^3)$. \\
When $\lambda_{*}>1$, the body starting with liminal velocity must immediately leave the ground and therefore fly-off phase dissapears (as starting with any $\lambda<\lambda_{*}$ leads to halting; see appendix \ref{app_behavior_of_Lambda}). The larger the friction coefficient $\mu$, the smaller $A$ is required for the fly-off forbidance to occur. It also should be noticed, that this effect limits the allowed $\theta$ angles to be $<\theta_*$.\\
Such phase disappearance is new and  surprising in comparison to the zero-drag situation.
\paragraph{Approaching the bottom of the hill}
The only way to approach the bottom of the hill is to reduce speed to zero in $\theta=\pi/2$. Since $\theta_*$ is the greatest halt angle accessible for a given set of $A$ and $\mu$ (also see appendix \ref{app_behavior_of_Lambda}), $\lambda_*$ corresponding to that angle corresponds to the minimal velocity, required for the equator approach. If we calculate the limit of $\lambda_{*}(\mu, A)$ from Eq. (\ref{lambda*}) as $\mu \rightarrow \infty$ ($\theta_{*} = \arctan\mu \rightarrow \pi/2$) we obtain:
\begin{equation}
	\lim\limits_{\mu \rightarrow \infty}\lambda_* = 1 \text{,}
\end{equation}
which happens to equal exactly one. Nevertheless, we have shown that for sufficiently large $\mu$ any non-zero drag $A>(3+ 12 \mu^2)/(-10\mu+8\mu^3)\rightarrow0$ will cause the liminal velocity value to exceed $1$ at some point, therefore making it impossible for the skier to achieve angle $\theta_{*}\rightarrow\pi/2$. \\
It turns out that the outcome obtained for a simpler case in Ref. \onlinecite{Kufel, Mungan} does not sustain for a situation with any nonzero drag force (no matter how small) and skier can never get to the bottom of the hill, even with the velocity infinitesimally close to the one causing the body to immediately fall off and an infinite friction coefficient.

\subsection{Exact solution approach} \label{Sect_ExactSol}
\begin{figure}
	\includegraphics[width=\columnwidth]{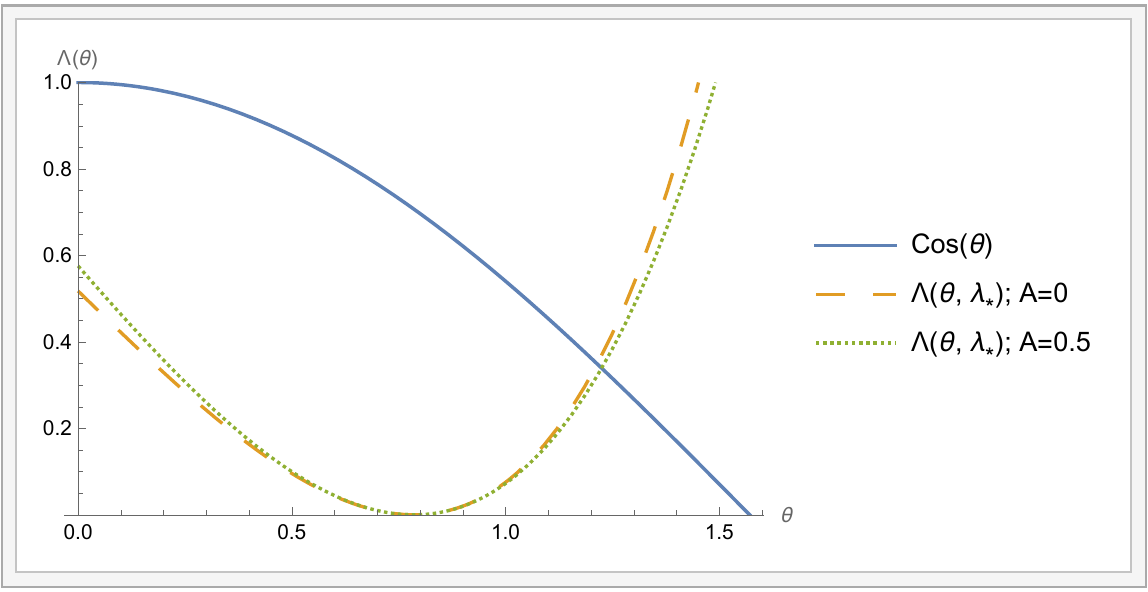}
	\caption{\justifying
		This figure shows liminal trajectories (given by inserting RHS of Eq.  (\ref{lambda*}) into Eq. (\ref{LambdaExactSolution})) for $\mu=1$ for a model with and without drag force. We can see, that including small drag force does not influence the existence of discontinuous phase transition effect much, as it almost does not change the liminal curve itself.}
	\label{fig2}
\end{figure}

\paragraph{Finding the liminal motion}
We substitute $\theta=\theta_{*}$ given by Eq. (\ref{theta*}) into Eq. (\ref{LambdaExactSolution}) and - having compared the left side to zero - solve for $\lambda_*$:
\begin{equation}\label{lambda_*_exact}
	\lambda_*=\frac{2 \sqrt{\mu ^2+1} e^{(-2\mu + A ) \arctan(\mu )}+2 \mu  (2 \mu - A )-2}{(2 \mu -A )^2+1} \text{.}
\end{equation}
\paragraph{Influence on the phase-boundary}
First, let's consider an impact of adding drag force to the problem on the liminal curve itself and thus the discontinuous phase transition effect. To calculate the derivative of $\Lambda(\lambda_*)$ with respect to $A$ we must take into account $\lambda_*$ also being a function of $A$.
\begin{equation}
	\frac{d\Lambda}{d A} = \frac{\partial \Lambda (\theta ,A,\mu ,\lambda_{*} )}{\partial A}+\frac{d \lambda _*(A,\mu )}{d A} \frac{\partial \Lambda (\theta ,A,\mu ,\lambda_* )}{\partial \lambda_*} \text{.}
\end{equation}
Since we are interested in how adding drag force to an idealized non-drag situation influences the model, we evaluate the above at $A=0$ and get:
\begin{eqnarray}
	\frac{d\Lambda}{d A} &=& e^{2\mu(\theta-\arctan\mu)}\left(\frac{2\sqrt{\mu^2+1}}{4\mu^2+1} (\theta-\arctan\mu) \right. \nonumber\\	 
	&&\left. - \frac{8\mu\sqrt{\mu^2+1}}{(4\mu^2+1)^2} \right) + \sin\theta \left(\frac{16\mu^2-2}{(4\mu^2+1)^2} \right) \nonumber \\ 
	&& + \cos\theta \left( \frac{10\mu-8\mu^3}{(4\mu^2+1)^2}\right) \text{.}
\end{eqnarray}		
Numerical analysis shows, that this expression ($\theta \in \left[0, \frac{\pi}{2}\right]$) takes only small values (approx. from the interval $\left[-0.16,~1.2\right]$). It means, that even if for certain conditions adding drag force can violently influence curves very near the liminal one, as has been shown in the foregoing paragraph and in Fig. \ref{fig1}, it almost does not impact the phase-boundary itself, and thus the occurrence of the discontinuous phase transition effect. It can be seen in both Figs. \ref{fig1} and \ref{fig2}. 

\paragraph{Motion nearby phase-boundary}

It is interesting to observe, how minor changes in initial velocity can change the angle at which the movement finishes. To get some quantitative conclusions, we can differentiate $\Lambda$ with respect to $\lambda$:
\begin{equation}\label{10}
	\frac{d\Lambda}{d\lambda}=e^{\theta(2\mu-A)} \text{.}
\end{equation}
We can see that this derivative grows exponentially with angle $\theta$. Therefore, because trajectories that are beginning with $\lambda$ different by a small value from $\lambda_*$ are corresponding to flying off or halting at the largest angles possible, final angles corresponding to these trajectories are at the greatest risk of being seriously shifted by a very small change of $\lambda$ (as it is shown in the Fig. (\ref{fig1})). The greater the values of $\mu$ and the smaller the value of $A$, the stronger will be that shift. Moreover, if the final angle ($\lambda$ is near $\lambda_*$) and $\mu$ is sufficiently high, even minimal change in $A$ (such as adding a small drag force to the idealized model without it) can deepen that shift in an arbitrarily major way and even lead to the phase transition.\\
Furthermore, because we have shown, that the liminal curve $\Lambda(\theta,\lambda_*,A)$ is not very vulnerable to 	changes in $A$, we can approximate $\Lambda(A,\lambda_*) \approx \Lambda(0,\lambda_*)$. From Eq. (\ref{10}), for some small $\epsilon$, we can write:
\begin{eqnarray}
	\Lambda(A,\lambda_*+\epsilon) &\approx& \Lambda(A,\lambda_*) + \epsilon \frac{d \Lambda}{d \lambda} \nonumber\\
	&\approx& \Lambda(0,\lambda_*) + \epsilon e^{2\mu\theta} e^{-A\theta} \nonumber \\
	&\approx& \Lambda(0,\lambda_*) + \epsilon e^{2\mu\theta} - \epsilon A \theta e^{2\mu\theta} + \dots \nonumber \\
	&\approx& \Lambda(0,\lambda_*+\epsilon) - \epsilon A \theta e^{2\mu\theta} + \dots ~ \text{.}
	\label{61}
\end{eqnarray}
Now we can read out, that nearly liminal trajectories can be strongly influenced by adding drag force to the problem, provided that $\mu$ and $\theta$ are sufficiently high. Once again, it surprisingly appears, that as $\mu$ grows, so do the drag-induced changes.  This agrees with an outcome obtained with the perturbative approach earlier.
\paragraph{Fly-off phase forbiddance}
Analysis of Eqn. (\ref{lambda_*_exact}) also confirms fly-off phase disappearance, that occurs when $\lambda_{*}(A,\mu)>1$. It may be noticed, that for $A = 2\mu$, Eqn. (\ref{lambda_*_exact}) takes a surprisingly simple form:
\begin{equation}
	\lambda_{*} = 2\sqrt{1+\mu^2}-2 \text{.}
\end{equation}
In this case, phase forbidance effect arises if $\mu= A/2 >\sqrt{5}/2$.
\section{Discussion}
In this article, I have developed a solution to an extension of the well-known undergraduate mechanics problem of the skier on the hemispherical hill with friction and drag force. I have utilized exact analytical and complimentary perturbative methods to develop insight into the problem. This enabled acquiring some quantitative results on the robustness of the previously known solutions to drag force inclusion.\\
The most important aspect of this study is an observation and explanation of the counter-intuitive characteristics of the system. It has been proved for certain values of the initial speed $\lambda \rightarrow \lambda_{\text{Amp}}(\mu)$, that omitting drag force will lead to significant first-order inaccuracies in the system description. This leads to an unintuitive conclusion: even for very large friction, there exists such a skier's initial velocity, that the motion will be highly affected by adding the drag force to the model without it. Moreover, it is shown, that the larger the friction coefficient $\mu$, the more visible the differences  will be.\\
Furthermore, the impact of adding drag on the phase transition between two possible ending states of the motion, is being analyzed. The initial velocity separating them was found in an exact and perturbative form. Interestingly, it has been shown, that for certain values of friction and drag coefficients, the only possible outcomes of the skier's motion is halting - for any value of its initial velocity $\lambda$, the fly-off phase disappears. \\
The paper holds didactic value as it shows an application of the perturbation theory -- best known from quantum mechanics -- in a simple classical problem.

%% file: appendices.tex
\section{Solving reccurence relation for $\beta_{1}^n$ and $\beta_{2}^n$} \label{app_solving_reccurence_relation}
In order to solve the system of recurrence relations (\ref{beta1_rec}-\ref{beta2_rec}), we want to find generating functions $B_{1}(x)$ and $B_{2}(x)$ of series $\beta_{1}^{n}$ and $\beta_{2}^{n}$ respectively (for more details about generating functions themselves see Ref. \onlinecite{GeneratingFunctions}). Therefore, we multiply both sides of equations (\ref{beta1_rec}-\ref{beta2_rec}) by $x^{n+1}$ and sum for all $n$ from 0 to infinity:
\begin{eqnarray}
	\sum_{n\geq 0}\beta_{1}^{n+1}x^{n+1} &=& \frac{1}{1+4\mu^2}\left(2\mu\sum_{n\geq 0}\beta_{1}^{n}x^{n+1}\right. \nonumber \\ 
	&& \left. -\sum_{n\geq 0}\beta_{2}^{n}x^{n+1}\right) \text{;} \\
	\sum_{n\geq 0}\beta_{2}^{n+1}x^{n+1} &=& \frac{1}{1+4\mu^2}\left(2\mu\sum_{n\geq 0}\beta_{2}^{n}x^{n+1} \right. \nonumber \\
	&& \left.+\sum_{n\geq 0}\beta_{1}^{n}x^{n+1}\right) \text{.}
\end{eqnarray}
From the generating function definition we obtain:
\begin{eqnarray}
	B_{1}(x) - \beta_{1}^{0} &=& \frac{x}{1+4\mu^2}\left(2\mu  B_{1}(x)- B_{2}(x)\right) \text{;} \\
	B_{2}(x) - \beta_{2}^{0} &=& \frac{x}{1+4\mu^2}\left(2\mu  B_{2}(x)+ B_{1}(x)\right) \text{.}
\end{eqnarray}
This system of equations can be easily solved for $B_{1}(x)$ and $B_{2}(x)$:
\begin{eqnarray}
	B_{1}(x) &=& \frac{\beta_{1}^{0} \left(4 \mu ^2+1\right)-x (2 \beta_{1}^{0} \mu +\beta_{2}^{0})}{(2 \mu +x)^2+1} \text{;} \label{B1_closed}\\
	B_{2}(x) &=& \frac{\beta_{2}^{0} \left(4 \mu ^2+1\right)+x (-2 \beta_{2}^{0} \mu +\beta_{1}^{0})}{(2 \mu +x)^2+1} \text{.}
\end{eqnarray}
Now we want to express these generating functions in an open form of $\sum_{n\geq 0}\beta_{i}^{n}x^{n}$ to read out the $\beta_{i}^n$. To do these we opt to show them in terms of infinite sums of geometric series of a well-known form $a_{0}/(1-qx)$. We shall perform further manipulations just for $B_{1}(x)$ and then obtain $B_{2}(x)$ by noticing certain symmetries in the original set of equations.\\
First, we shall factor the denominator over complex numbers (since it does not have real roots) and disassemble the fraction as:
\begin{equation}
	B_{1}(x) = \frac{1}{2} \left(\frac{\beta _1^0+i \beta _2^0}{1+\frac{x}{-2 \mu +i}}+\frac{\beta _1^0-i \beta _2^0}{1+\frac{x}{-2 \mu -i}}\right) \text{.}
\end{equation}
From the properties of geometric series, we rewrite it in the desired form:
\begin{eqnarray}
	B_{1}(x)
	&=& \frac{1}{2}\sum_{n\geq0}\left[\left(\frac{1}{-2 \mu + i} \right)^{n}\left(\beta_{1}^{0} + i \beta_{2}^{0} \right) \right. \nonumber \\
	&& \left. + \left(\frac{1}{-2 \mu - i} \right)^{n}\left(\beta_{1}^{0} - i \beta_{2}^{0} \right) \right] \text{.}
\end{eqnarray}
The $n$-th order coefficient in this series is simply $\beta_{1}^{n}$, so:
\begin{eqnarray}
	\beta_{1}^{n} &=& \frac{1}{2}\left[\left(\frac{1}{-2 \mu + i} \right)^{n}\left(\beta_{1}^{0} + i \beta_{2}^{0} \right) + \left(\frac{1}{-2 \mu - i} \right)^{n}\left(\beta_{1}^{0} - i \beta_{2}^{0} \right) \right] \nonumber \\
	&=& \mathfrak{Re} \left(\left(\frac{1}{-2 \mu + i} \right)^{n}\left(\beta_{1}^{0} + i \beta_{2}^{0} \right) \right) \text{.} \label{beta1_rec_App}
\end{eqnarray}
It is worth noticing that this number must be real, as it is a sum of a complex number and its complex conjugate. It is a reasonable outcome because $\beta_{1}^{n}$ was originally defined by the recurrent relation involving solely real numbers.\\
Equation (\ref{beta1_rec}) transforms into Eq. (\ref{beta2_rec}) by changing $\beta$ indexes from 1 to 2 and 2 to 1 and after that by changing $\beta_{1}^{n}$ into $-\beta_{1}^{n}$. We can use that symmetry to obtain an expression for $\beta_{2}^{n}$ from (\ref{beta1_rec_App}):
\begin{eqnarray}
	\beta_{2}^{n} &=& \frac{1}{2}\left[\left(\frac{1}{-2 \mu + i} \right)^{n}\left(\beta_{2}^{0} - i \beta_{1}^{0} \right) + \left(\frac{1}{-2 \mu - i} \right)^{n}\left(\beta_{2}^{0} + i \beta_{1}^{0} \right) \right] \nonumber \\
	&=& \mathfrak{Re} \left(\left(\frac{1}{-2 \mu + i} \right)^{n}\left(\beta_{2}^{0} - i \beta_{1}^{0} \right) \right) \text{.}
\end{eqnarray}
\section{Behavior of $\Lambda(\theta)$ function} \label{app_behavior_of_Lambda}

We want show, that the body cannot fly off for $\lambda<\lambda_{*}$ or halt for $\lambda>\lambda_{*}$ and that $\theta_{*}$ is the biggest angle at which it can halt. With this aim in mind, we generalize method that was used in Ref. \onlinecite{Kufel} where authors studied function $\Lambda(\theta,\lambda,\mu,A=0)$, which apart from adjusting to non-zero drag case we also modify, so that it can be used without refering to exact form of $\Lambda(\theta)$ function (\ref{LambdaExactSolution}). \\ 
We design a function $\lambda_{\#}(\theta_{\#},\mu,A)$, such that $d\Lambda/d\theta(\lambda_{\#}(\theta_{\#},\mu,A),\theta_{\#},\mu,A) = 0$. It will represent an initital velocity, that would cause $\Lambda(\theta)$ to flatten out at a given angle $\theta_{\#}$. This condition and Eq. (\ref{DiffEqua_Lambda}) imply equality (for $2\mu =A$ we define $\Lambda(\lambda_{\#},\mu,A) = 0$, since in that case from Eq. (\ref{DiffEqua_Lambda}) we always have $\theta_\# = \theta_{*}$ and by definition $\lambda_{\#}=\lambda_{*}$):
\begin{equation}
	\Lambda(\lambda_{\#}(\theta_\#,\mu, A),\theta_{\#},\mu,A) = \frac{2\mu \cos\theta_\# - 2 \sin\theta_\#}{2\mu-A} \label{Lambda_lambdahash} \text{.}
\end{equation}
Using (\ref{DiffEqua_Lambda}) we calculate the second derivative of $\Lambda(\theta)$ with respect to $\theta$ at point $\theta=\theta_{\#}$:
\begin{equation}
	\left.\frac{d^2\Lambda(\lambda_{\#}(\theta_{\#},\mu,A),\theta,\mu,A)}{d\theta^2}\right|_{\theta=\theta_{\#}} = 2 \cos\theta_\# + 2\mu\sin\theta_{\#} > 0 \text{,}
\end{equation}
which is positive for any angle $\theta_\# \in [0,\pi/2)$. Therefore,  in this interval any function $\Lambda(\theta)$ always must be concave up and consequently can have only one extremum in it.\\
From the $\Lambda(\theta,\lambda,A,\mu)$ monotonicity in the $\lambda$ direction (see Eqns. (\ref{LambdaExactSolution}) or (\ref{LambdaExpansion})/(\ref{bn_gen})) we obtain two equivalent conditions (For $\epsilon>0$):
\begin{eqnarray}
	\Lambda(\theta,\lambda-\epsilon,A,\mu) &<& \Lambda(\theta,\lambda,A,\mu) \label{LambdaMonoCond_1} \text{;}\\
	\Lambda(\theta,\lambda,A,\mu) &<& \Lambda(\theta,\lambda+\epsilon,A,\mu) \label{LambdaMonoCond_2} \text{.}
\end{eqnarray}
We know that $\theta=\theta_*$ and $\lambda_*$ are defined in such a way, that $\Lambda(\theta_*,\lambda_*, A, \mu)=0$ is true. \\
Let us now consider motion, beginning with velocity $\lambda_*-\epsilon$. From Eq. (\ref{LambdaMonoCond_1}) we have:
\begin{eqnarray}
	\Lambda(\theta_*,\lambda_*-\epsilon,A,\mu) &<& \Lambda(\theta_*,\lambda_*,A,\mu) = 0  \text{,}\label{B_cond_1}
\end{eqnarray}
and from the definition of initial velocity:
\begin{equation}
	\Lambda(0,\lambda_*-\epsilon,A,\mu) = \lambda_*-\epsilon >0 \label{B_cond_2} \text{.}
\end{equation}
Now, thanks to Eqs. (\ref{B_cond_1}) and (\ref{B_cond_2}) and since $\Lambda(\theta)$ is continuous, we can apply Darboux theorem to conclude that for each $\epsilon<\lambda_*$, there exists such $\theta_{\text{halt}} \in \left(0,~ \theta_* \right) $, that satisfies the condition $\Lambda(\theta_{\text{halt}},\lambda_*-\epsilon,A,\mu)=0$. Therefore, for a given set of $(A, \mu)$, beginning motion with $\lambda<\lambda_*$ corresponds to halting in an angle $\theta_{\text{halt}}<\theta_{*}$.\\
On the other hand for $\lambda_*$ an angle $\theta=\theta'\leq \pi/2$ exists ($\Lambda(\theta, \lambda_{*})$ is concave up and has local minimum $\Lambda(\theta_{*})=0$ for $\theta=\theta_{*}<\pi/2$), such that $\Lambda(\theta',\lambda_*,A,\mu)= \cos\theta'$. It is an angle of flying off, having started motion with velocity infinitisimally greater than $\lambda_*$. Let us consider a motion, that begins with velocity $\lambda_*+\epsilon$. From Eq. (\ref{LambdaMonoCond_2}) we have:
\begin{gather}
	\Lambda(\theta',\lambda_*+\epsilon,A,\mu) > \Lambda(\theta',\lambda_*,A,\mu)=\cos\theta' \nonumber \text{,} \\
	\Lambda(\theta',\lambda_*+\epsilon,A,\mu) - \cos\theta' > 0 \text{.} \label{B_cond_3}
\end{gather}
and once again from the definition of initial velocity:
\begin{gather}
	\Lambda(0,\lambda_*+\epsilon,A,\mu) = \lambda_*+\epsilon < \cos(0) \text{,} \nonumber \\
	\Lambda(0,\lambda_*+\epsilon,A,\mu) -\cos(0) < 0 \label{B_cond_4} \text{.}
\end{gather}
Since $\cos\theta$ and $\Lambda(\theta)$ are both continuous and because of Eqs. (\ref{B_cond_3}) and (\ref{B_cond_4}), on the basis of Darboux theorem we deduct, that for any $\lambda_*+\epsilon$ there exists $\theta=\theta_{\text{fly}} \in \left(0,~ \theta'\right)$, such that $\Lambda(\theta_\text{fly},\lambda_*+\epsilon,A,\mu) =\cos(\theta_\text{fly})$ - a point of intersection of these two curves and therefore fly-off angle, that is smaller than the angle corresponding to the beginning velocity equal $\lambda_*$. So we have showed, that for a given set of $(A, \mu)$, motion that has started with $\lambda>\lambda_*$ ends for an angle smaller than $\theta'$.\\
Because $\Lambda(\theta)$ is always concave up, has only one local minimum in the interval $[0, \pi/2)$ and satisfies conditions (\ref{LambdaMonoCond_1}-\ref{LambdaMonoCond_2}) the body cannot fly-off for $\lambda<\lambda_{*}$ or halt for $\lambda\geq \lambda_{*}$ as in those situations $\Lambda(\theta, \lambda)$ curve would have to intersect $\Lambda(\theta,\lambda_{*})$ on the $(\theta, \Lambda)$ plane. Our proof is therefore complete. $\square$
\section{Equivalence of exact and perturbative formulae for $\Lambda(\theta)$} \label{app_pertu_to_exact}
First, we explicitly write out perturbative formula for $\Lambda(\theta)$ according to Eqs. (\ref{LambdaExpansion}), (\ref{bn_gen_c_alt}) and (\ref{beta2_c}):
	\begin{eqnarray}
		\Lambda(\theta) &=& \lambda e^{2\mu\theta} \sum_{n=0}^{\infty}\frac{(-A\theta)^n}{n!}  \nonumber\\ 
		&& +  \mathfrak{Re}\left(\sum_{n=0}^{\infty} A^n \widetilde{\beta_2^{n}}\left(e^{i\theta}-e^{2\mu\theta}\sum_{k=0}^{n}\frac{(i-2\mu)^k\theta^k}{k!}\right)\right) \nonumber\\
		&=& \lambda e^{2\mu\theta} \sum_{n=0}^{\infty}\frac{(-A\theta)^n}{n!} \nonumber \\
		&& + \mathfrak{Re}\left(\sum_{n=0}^{\infty} 2 A^n \frac{i+\mu}{(2\mu-i)^{n+1}}\left(e^{i\theta} \right. \right. \nonumber \\
		&& \qquad \left. \left. - e^{2\mu\theta}\sum_{k=0}^{n}\frac{(i-2\mu)^k\theta^k}{k!}\right)\right)\text{.}
	\end{eqnarray}
Then, using power series form of exponential function together with formula for the geometric series sums we perform the following algebraic manipulations:
\begin{widetext}
	\begin{eqnarray}
		\Lambda(\theta)
		&=& 
		\lambda e^{\theta(2\mu- A)} + \mathfrak{Re}\left(2 \frac{i+\mu}{2\mu-i} e^{i\theta} \sum_{n=0}^{\infty}\left(\frac{A}{2\mu-i}\right)^n \right) \nonumber\\
		&& \qquad- \mathfrak{Re}\left(2 \frac{i+\mu}{2\mu-i} e^{2\mu\theta}\sum_{n=0}^{\infty}\sum_{k=0}^{n}\left(\frac{A}{2\mu-i}\right)^n \frac{(i-2\mu)^k\theta^k}{k!}\right) \nonumber\\
		&=& 
		\lambda e^{\theta(2\mu- A)} + \mathfrak{Re}\left(2 \frac{i+\mu}{2\mu-i} \cdot \frac{e^{i\theta}}{1- \frac{A}{2\mu-i}}\right) \nonumber\\
		&& \qquad -  \mathfrak{Re}\left(2 \frac{i+\mu}{2\mu-i} e^{2\mu\theta}\sum_{k=0}^{\infty} \frac{(i-2\mu)^k\theta^k}{k!}\sum_{n=k}^{\infty}\left(\frac{A}{2\mu-i}\right)^n \right) \nonumber \\
		&=& 
		\lambda e^{\theta(2\mu- A)} + \mathfrak{Re}\left(\frac{2(i+\mu)}{2\mu-i-A}e^{i\theta}\right)\nonumber \\
		&& \qquad- \mathfrak{Re}\left(2 \frac{i+\mu}{2\mu-i} e^{2\mu\theta}\sum_{k=0}^{\infty} \frac{(i-2\mu)^k\theta^k}{k!} \left(\sum_{n=0}^{\infty}\left(\frac{A}{2\mu-i}\right)^n - \sum_{n=0}^{k-1}\left(\frac{A}{2\mu-i}\right)^n \right)\right) \nonumber \\
		&=& 
		\lambda e^{\theta(2\mu- A)} + \mathfrak{Re}\left(\frac{2(i+\mu)}{2\mu-i-A}e^{i\theta}\right)- \mathfrak{Re}\left(2 \frac{i+\mu}{2\mu-i}e^{2\mu\theta}\sum_{k=0}^{\infty} \frac{(i-2\mu)^k\theta^k}{k!} \cdot \frac{\left(\frac{A}{2\mu-i}\right)^k}{1- \frac{A}{2\mu-i}}\right) \nonumber \\
		&=& 
		\lambda e^{\theta(2\mu- A)} + \mathfrak{Re}\left(\frac{2(i+\mu)}{2\mu-i-A}e^{i\theta}\right)- \mathfrak{Re}\left(\frac{2(i+\mu)}{2\mu-i-A}e^{2\mu\theta}\sum_{k=0}^{\infty} \frac{(-A\theta)^k}{k!}\right) \nonumber\\
		&=& 
		\lambda e^{\theta(2\mu- A)} + \mathfrak{Re}\left(\frac{2(i+\mu)}{2\mu-i-A}e^{i\theta}\right)- \mathfrak{Re}\left(\frac{2(i+\mu)}{2\mu-i-A}\right)e^{\theta(2\mu-A)} \text{.} \label{App_D_Lambda_theta_pert}
	\end{eqnarray}
\end{widetext}
Now, by calculating the real parts of the appropriate complex numbers, the reader can easily verify the equivalence of RHS's of Eqs. (\ref{App_D_Lambda_theta_pert}) and (\ref{LambdaExactSolution}). Since all of the manipulations performed would be equally valid if reversed, we have therefore  proofed the equivalence between exact and perturbative formulae for $\Lambda(\theta)$. $\square$